\documentclass{aa}  
\usepackage{graphicx}
\usepackage{txfonts}
\begin{document}
   \title{Searching for spiral features in the outer Galactic disk.}
   \subtitle{The field towards WR38 and WR38a}

   \author{Giovanni Carraro
          \inst{1,2}
          \and
           Edgardo Costa\inst{3}
          }

   \offprints{G. Carraro}

   \institute{
             $^1$ESO, Casilla 19001, Santiago, Chile\\
             $^2$Dipartimento di Astronomia, Universit\`a di Padova,
                 Vicolo Osservatorio 2, I-35122 Padova, Italy\\
             $^3$Departamento de Astronom\'ia, Universidad de Chile,
                 Casilla 36-D, Santiago, Chile\\
                 }

   \date{Received ....; accepted...}

  \abstract
   {The detailed spiral structure in the outer Galactic disk is still poorly known, and for
   several Galactic directions we still have to rely on model extrapolations.}
   {One of these regions is the fourth Galactic quadrant, in the sector comprised between
   Vela and Carina ($270^o \leq l \leq 300^o$) where -apart from the conspicuous Carina
   branch of the Carina Sagittarius arm- no spiral arms have been detected so far in the optical
   beyond $l \sim 270^o$.}
   {By means of deep $UBVI$ photometry, we search for spiral features in known low absorption
   windows.  $U$ photometry, although observationally demanding, constitutes a powerful
   tool to detect and characterize distant aggregates of young stars, and allows to derive
   firmer distance estimates. We have studied a direction close to the tangent ($l \sim 290^o$)
   to the Carina arm, in an attempt to detect optical spiral tracers far beyond the Carina branch,
   where radio observations and models predictions seem to indicate the presence of the extension
   of the Perseus and Norma-Cygnus spiral arms in the fourth quadrant.}
   {Along this line of sight, we detect three distinct groups of young stars. Two of them, at
   distances of $\sim$2.5 and $\sim$6.0 kpc, belong to the Carina spiral arm (which is crossed twice
   in this particular direction). Interestingly, the latter is here detected for the first time.
   The third group, at a distance of $\sim$12.7 kpc, is likely a
   part of the Perseus arm which lies beyond the Carina arm, and constitutes the first optical
   detection of this arm in the fourth Galactic quadrant. The position of this feature is compatible
   both with HI observations and model predictions. We furthermore present evidence that this
   extremely distant group, formerly thought to be a star cluster (Shorlin~1), is in fact a diffuse
   young population typical of spiral features. In addition,  our data-set does not support, 
   as claimed in the
   literature,  the possible presence of the Monoceros Ring toward this direction}
   {This study highlights how multicolor optical studies can be effective to probe the spiral structure
   in the outer Galactic disk. More fields need to be studied in this region of the Galaxy to better
   constrain the spiral structure in the fourth Galactic quadrant, in particular the shape and extent
   of the Perseus arm, and, possibly, to detect the even more distant Norma-Cygnus arm.}


   \maketitle
%

\section{Introduction}

\noindent
In recent times the study of the spiral structure of the outer
Galactic disk has witnessed a renewed interest 
(see e.g. Levine et al. 2006, V\'azquez et al. 2008, Benjamin 2008),
in part due to the claimed discovery of structures in the form of star
over-densities which would be produced by accretion/merging events.
Examples are those in Monoceros (Yanny et al. 2003), Canis Major (Martin
et al. 2004) and Argus (Rocha Pinto et al. 2006). To assess the reality
and properties of these over-densities, a detailed investigation of the
outer Galactic disk is a basic requirement. Such an investigation would
also improve our knowledge of the extreme periphery of the Milky Way (MW).\\

\noindent
Spiral features can be detected by using a variety of tracers, namely
HI (see e.g. Levine et al. 2006), HII (see e.g. Russeil 2003), CO (see e.g.
Luna et al. 2006), and optical objects (see e.g. V\'azquez et al 2008).
These diverse techniques have provided a coherent  picture of the spiral
structure in the third Galactic quadrant (Moitinho et al. 2008). Our group
has contributed to this effort providing optical information for a large
sample of young open clusters (Moitinho et al. 2006), in the field of which
we have recognized distant and reddened sequences which allowed us to define
the shape and extent of spiral arms up to 20 kpc from the Galactic center.\\

   \begin{figure*}
   \centering
   \includegraphics[width=18.5cm]{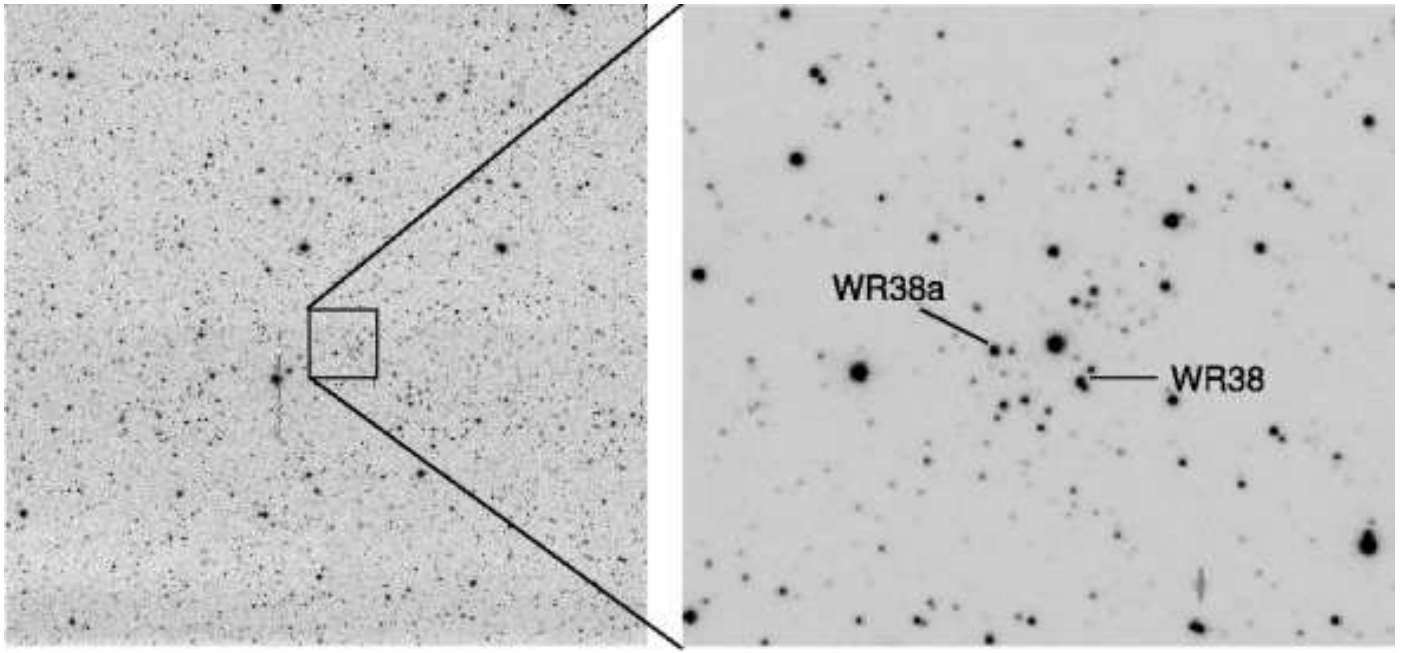}
   \caption{Left panel: $V$-band 100 sec exposure image of the field observed
    in the area of the star cluster Shorlin~1.  Field shown is $20\arcmin$ on a side.
    North is up and East to the left. Right panel: Zoom of the cluster region,
    in which the two WR stars (WR38 and WR38a) are indicated.  Field shown is
    $\sim3\arcmin$ on a side.}%
    \end{figure*}
\noindent
This paper is the first result of a study of selected low-absorption regions in
the fourth Galactic quadrant, aimed at finding distant optical spiral features.
We present an investigation of the field towards the Wolf-Rayet (WR) stars WR38 and
WR38a ($RA = 11^{h}06^{m}$, $DEC = -61^{o}14^{\prime}$, Shorlin et al. 2004, hereafter Sh04; Wallace et al. 2005, hereafter Wa05),
which attracted our attention for several reasons.\\

Sh04 and Wa05 discuss the discovery of a distant, young, compact
star cluster, at a distance of $\sim$12 kpc, associated with these two WR stars.
These two studies, however, reach quite different conclusions about the relationship
of this cluster with known spiral arms.  While Sh04 propose that the cluster
is associated with an extension of the Perseus arm beyond the Carina-Sagittarius arm,
Wa05, more conservatively, suggest that it lies in the more distant part of the Carina
arm. In fact, the line-of-sight in the direction to WR38 and WR38a crosses twice the
Carina-Sagittarius arm. In a quite different interpretation, Frinchaboy et al. (2004)
associate this cluster to the Galactic Anticenter Stellar Structure (GASS), a
population of old star clusters, possibly members of the Monoceros Ring.\\

With the purpose of clarifying these issues, we undertook an observational campaign
which addresses the following questions: Is the group of stars close to WR38 and WR38a
(i.e. Shorlin 1) a real star cluster?  Does this group belong to the most distant
part of the Carina arm or to the Perseus arm?  Are we looking in this direction at
signatures of the Monoceros ring, or the Argus system?\\

\noindent
The paper is organized as follows. In Sect~2 we describe the observations and the
reduction procedure and compare our data-set with previous investigations,
and in Sect.~3 we discuss the reddening law in the Galactic direction under consideration.
In Sect.~4 and 5 we thoroughly discuss the properties of the three different stellar
populations detected in the field observed. Finally, Sec.~6 summarizes the results of
our study.

\section{Observations and Data Reduction}

\subsection{Observations}

The region of interest (see Fig.~1) was observed with the Y4KCAM camera
attached to the 1.0m telescope, which is operated by the SMARTS consortium\footnote{{\tt http://http://www.astro.yale.edu/smarts}}
and located  at Cerro Tololo Inter-American Observatory (CTIO). This camera
is equipped with an STA 4064$\times$4064 CCD with 15-$\mu$ pixels, yielding a scale of
0.289$^{\prime\prime}$/pixel and a field-of-view (FOV) of $20^{\prime} \times 20^{\prime}$ at the
Cassegrain focus of the CTIO 1.0m telescope. The CCD was operated without binning, at a nominal
gain of 1.44 e$^-$/ADU, implying a readout noise of 7~e$^-$ per quadrant (this detector is read
by means of four different amplifiers). QE and other detector characteristics can be found at:
http://www.astronomy.ohio-state.edu/Y4KCam/detector.html.\\

\noindent
The observational material was obtained in three observing runs, resumed in Table~1. In our
first run we took deep $UBVI$ images of Shorlin~1, under good seeing, but non-photometric
conditions. In our second run we took medium and short exposures of Shorlin~1, and
observed Landolt's SA~98 $UBVRI$ standard stars area (Landolt 1992), to tie our $UBVI$
instrumental system to the standard system.  In a final third run we secured an additional
set of $U$-band images of Shorlin~1. Average seeing was 1.2$\arcsec$.\\

\begin{table}
\tabcolsep 0.1truecm
\caption{Log of $UBVI$ photometric observations.}
\begin{tabular}{lcccc}
\hline
\noalign{\smallskip}
Target& Date & Filter & Exposure (sec) & airmass\\
\noalign{\smallskip}
\hline
\noalign{\smallskip}
Shorlin~1 & 24 May 2006     & U & 60, 1500             &1.28$-$1.33\\
          &                 & B & 30, 100, 1200        &1.20$-$1.35\\
          &                 & V & 30, 100, 900         &1.22$-$1.34\\
          &                 & I & 100, 700             &1.25$-$1.34\\
Shorlin~1 & 29 January 2008 & U & 5, 15                &1.19$-$1.22\\
          &                 & B & 3, 5, 10             &1.19$-$1.23\\
          &                 & V & 3, 5, 10             &1.18$-$1.22\\
          &                 & I & 3, 5, 10             &1.18$-$1.23\\
SA~98     & 29 January 2008 & U & 2x10, 200, 300, 400  &1.17$-$1.89\\
          &                 & B & 2x10, 100, 2x200     &1.17$-$1.99\\
          &                 & V & 2x10, 25, 50, 2x100  &1.20$-$2.27\\
          &                 & I & 2x10, 50, 100, 2x150 &1.18$-$2.05\\
Shorlin~1 & 05 June 2008    & U & 30x3, 1800           &1.19$-$1.20\\
\noalign{\smallskip}
\hline
\end{tabular}
\end{table}

Our $UBVI$ instrumental photometric system was defined by the use of a standard broad-band
Kitt Peak $BVR_{kc}I_{kc}$ set in combination with a U+CuSO4 $U$-band filter. Transmission
curves for these filters can be found at: http://www.astronomy.ohio-state.edu/Y4KCam/filters.html.
To determine the transformation from our instrumental system to the standard Johnson-Kron-Cousins
system, we observed 46 stars in area SA~98 (Landolt 1992) multiple times, and with different
airmasses ranging from $\sim$1.2 to $\sim$2.3.  Field SA~98 is very advantageous, as it
includes a large number of well observed standard stars, and it is completely covered by
the CCD's FOV.  Furthermore, the standard's color coverage is very good, being:
$-0.5 \leq (U-B) \leq 2.2$; $-0.2 \leq (B-V) \leq 2.2$ and $-0.1 \leq (V-I) \leq 6.0$.\\

\subsection{Reductions}

Basic calibration of the CCD frames was done using the IRAF\footnote{IRAF is distributed
by the National Optical Astronomy Observatory, which is operated by the Association
of Universities for Research in Astronomy, Inc., under cooperative agreement with
the National Science Foundation.} package CCDRED. For this purpose, zero-exposure
frames and twilight sky flats were taken every night.  Photometry was then performed
using the IRAF DAOPHOT and PHOTCAL packages. Instrumental magnitudes were extracted
following the point spread function (PSF) method (Stetson 1987). A quadratic, spatially
variable, Master PSF (PENNY function) was adopted. The PSF photometry was finally
aperture-corrected, filter by filter. Aperture corrections were determined performing
aperture photometry of a suitable number (typically 10 to 20) of bright stars in the field.
These corrections were found to vary from 0.120 to 0.215 mag, depending on the filter.\\

\subsection{The photometry}

Our final photometric catalog consists of 7425 entries having $UBVI$ measures down to 
$V \sim $ 20, and 12250 entries having $BVI$ measures down to $V \sim 22$.\\

After removing saturated stars, and stars having only a few measurements in Landolt's (1992)
catalog, our photometric solution for a grand-total of 183 measurements in $U$ and $B$, and of 206
measurements in $V$ and $I$, turned out to be:\\

\noindent
$ U = u + (3.290\pm0.010) + (0.47\pm0.01) \times X - (0.010\pm0.016) \times (U-B)$ \\
$ B = b + (2.220\pm0.012) + (0.29\pm0.01) \times X - (0.131\pm0.012) \times (B-V)$ \\
$ V = v + (1.990\pm0.007) + (0.16\pm0.01) \times X + (0.021\pm0.007) \times (B-V)$ \\
$ I = i + (2.783\pm0.011) + (0.08\pm0.01) \times X + (0.043\pm0.008) \times (V-I)$ \\

The final {\it r.m.s} of the fitting was 0.073, 0.069, 0.035 and 0.030 in $U$, $B$, $V$ and $I$,
respectively.\\

\noindent
Global photometric errors were estimated using the scheme developed by Patat \& Carraro
(2001, Appendix A1), which takes into account the errors resulting from the PSF fitting
procedure (e.i. from ALLSTAR), and the calibration errors (corresponding to the zero point,
color terms and extinction errors). In Fig.~2 we present global photometric error trends 
plotted as a function of $V$ magnitude. Quick inspection shows that stars brighter than
$V \approx 20$ mag have errors lower than 0.10~mag in magnitude and lower than 0.20~mag in
all colors. \\

\begin{figure}s
   \centering
   \includegraphics[width=\columnwidth]{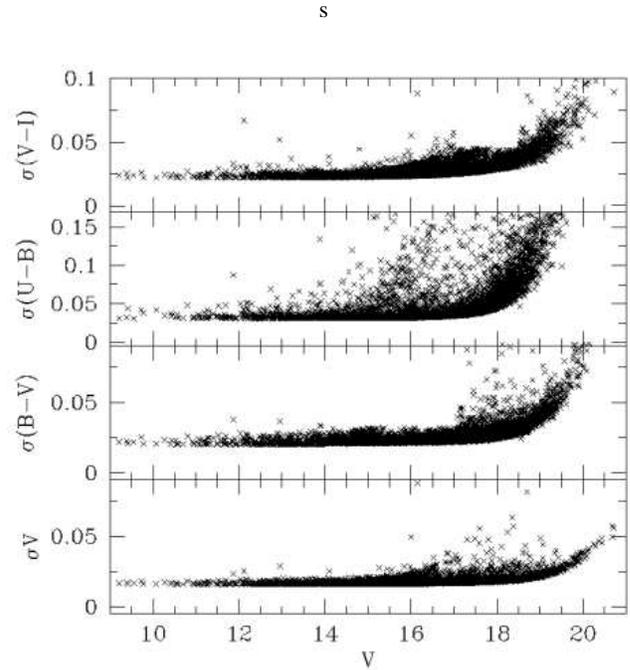}
   \caption{Photometric errors in V, (B-V), (U-B) and (V-I) as a function of V mag.}%
   \end{figure}

The only previous $UBV$ (Johnson-Morgan) photometric study of this region is that of Sh04,
who present photoelectric and CCD photometry for roughly 140 stars brighter than $V \sim$ 16.5
in the field of WR38 and WR38a.  In Fig.~3, we compare our photometry with that of Sh04 for
$V$, $B-V$ and $U-B$, in the sense ours minus theirs, as a function of our $V$ magnitude.
This comparison was possible only for 103 of their stars because, from the material presented
in Sh04, it was not possible to identify all objects measured by them (this issue was addressed
with D. Turner, private communication).  From the stars in common we obtain:
$\Delta V = 0.017 \pm 0.065$, $\Delta (B-V) = 0.014 \pm 0.066$ and $\Delta (U-B) = 0.034 \pm 0.199$.\\

\begin{figure}
   \centering
   \includegraphics[width=\columnwidth]{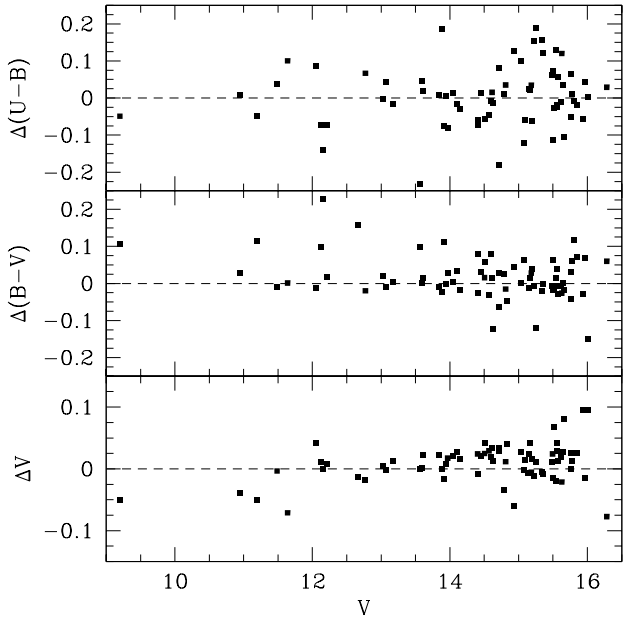}
   \caption{Comparison of our photometry with that of Sh04 for $V$, $B-V$ and $U-B$ (in the sense
    this work minus Sh04), as a function of our $V$ magnitude.}%
    \end{figure}

In spite of the significant scatter in the comparison for $U-B$, it can be concluded that the two
studies are in nice agreement. The poor match in this color may be ascribed to problems in their
$U-$band photometry, as extensively discussed in Sh04. Granted this, as a by-product we confirm that
the discrepancies between the photometry Sh04 and that of Wa05 is possibly due to the transformation
of the F336 ($U$), F439W ($B$), F555W ($V$) HST photometry to the Johnson-Morgan system.

\subsection{Astrometry}

For two-hundred stars in our photometric catalog there are J2000.0 equatorial coordinates
from the Guide Star Catalogue \footnote{Space telescope Science Institute, 2001, The Guide
Star Catalogue Version 2.2.02.}, version 2 (GSC-2.2, 2001). Using the SkyCat tool at ESO, and
the IRAF tasks ccxymatch and ccmap, we first established the transformation between our $(X,Y)$
pixel coordinates (from ALLSTAR) and the International Celestial Reference Frame (Arias et al.
1995). These transformations turned out to have an ${\it r.m.s.}$ value of 0.15$^{\prime\prime}$.
Finally, using the IRAF task cctran, we computed J2000.0 coordinates for all objects in our
catalog.

\section{The reddening law toward l=290.63 and b=-0.903}

A basic requirement before analyzing our photometric material is to investigate the reddening law
in the direction of WR38 and WR38a. Sh04 emphasize that according to their analysis the extinction
in this direction follows the normal law, namely that the ratio of total over selective absorption
$\frac{A_V}{E(B-V)}$ is 3.1. This value has already been found to be of general validity towards
Carina (Carraro 2002, Tapia et al. 2003).  The $(V-I)$ vs. $(B-V)$ color-color diagram show in Fig.~4.,
constructed using our $UBVI$ photometry, confirms the above statements. Most stars lie close to the
reddening-free Schmidt-Kaler (1982) Zero Age Main Sequence (ZAMS) line (dashed line), which runs
almost parallel to the reddening vector plotted in the upper-right corner of this figure. This 
vector has been drawn following the normal extinction ratio (Dean et al. 1978), which suggests that
in this specific Galactic direction absorption follows the normal law.  In the subsequent discussion
we will therefore adopt $\frac{A_V}{E(B-V)} = 3.1$ and $\frac{E(U-B)}{E(B-V)} = 0.72$.

\begin{figure}
   \centering
   \includegraphics[width=9.5truecm]{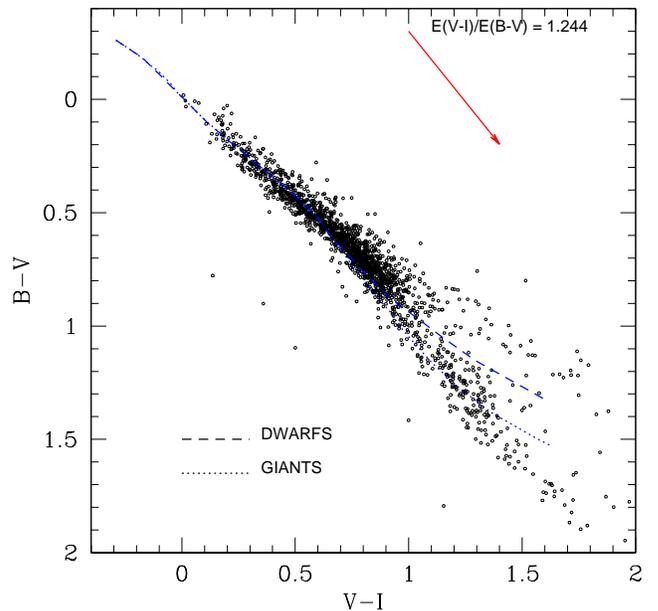}
   \caption{$(V-I)$ vs. $(B-V)$ color-color diagram for stars in our field with $UBVI$ photometry.
    The reddening vector for the normal extinction law is plotted in the upper-right corner. The
    the dashed and dotted lines represent the reddening-free Schmidt-Kaler (1982) Zero Age Main
    Sequence relations for dwarf and giant stars, respectively.}%
    \end{figure}

\section{Stellar populations in the field}

In Fig.~5. we present a $(B-V)$ vs. $(U-B)$ color-color diagram for those stars in our photometric
catalog which have photometric errors lower than 0.05 magnitudes in both $(B-V)$ and $(U-B)$.
This figure is similar to Fig.~3 of Sh04, except for the much larger number of stars resulting
from our larger area coverage and deeper photometry. The solid line is an empirical reddening-free
ZAMS for dwarf stars, from Schmidt-Kaler (1982). In agreement with what was discussed in the
previous section, we have adopted a normal reddening law for this region.  The corresponding
reddening vector has been plotted in the bottom of the figure.\\ 

\begin{figure*}
   \centering
   \includegraphics[width=15truecm]{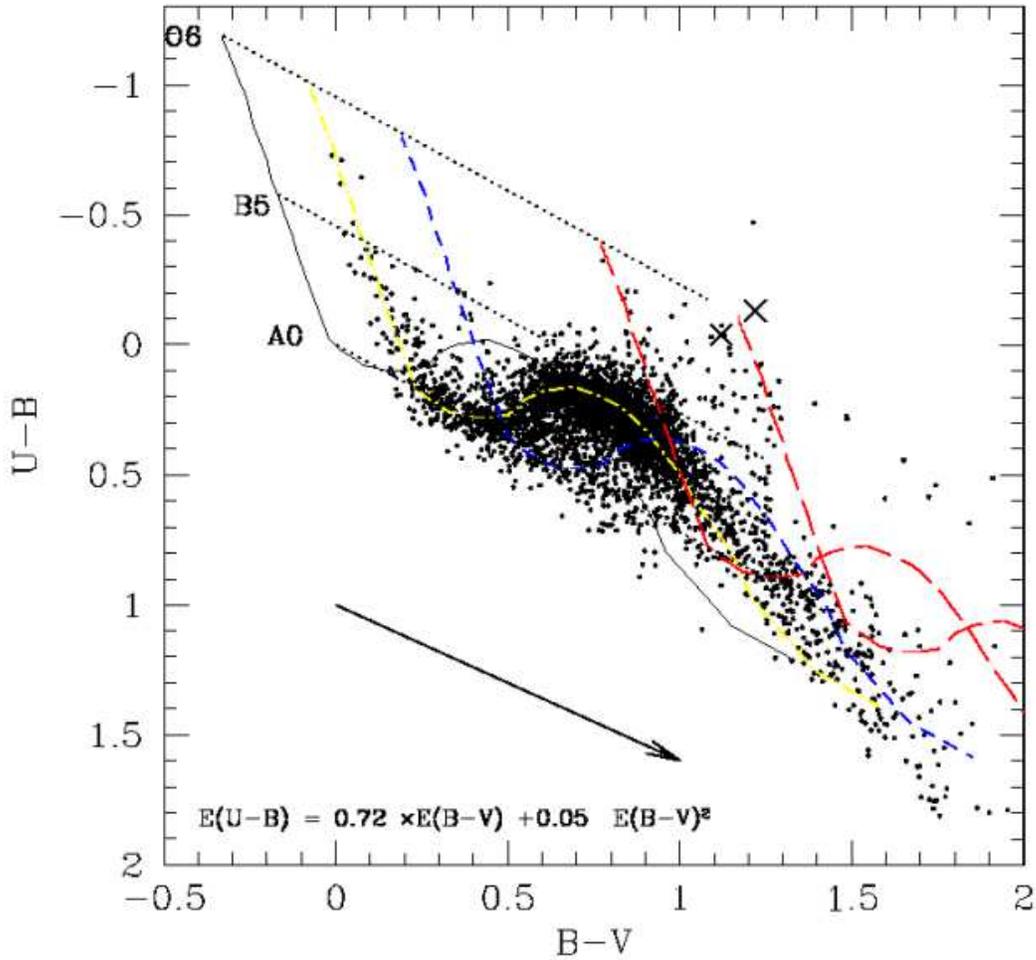}
   \caption{$(B-V)$ vs. $(U-B)$ color-color diagram for stars in our photometric catalog
    with photometric errors lower than 0.05 magnitudes in both $(B-V)$ and $(U-B)$. The
    reddening vector for the normal extinction law has been plotted in the bottom of the
    figure. The solid line is an empirical reddening-free ZAMS for dwarf stars, from
    Schmidt-Kaler (1982). The same ZAMS (dashed lines) has been displaced along the reddening
    line for three different amounts of reddening, 0.25, 0.52 and 1.45. The approximate
    location of stars with spectral type O6, B5 and A0 is also indicated. The two crosses are
    here used to show the position of WR38 and WR38a. See text for more details.}%
    \end{figure*}

\noindent
Shifting the ZAMS by different amounts in the direction of the reddening vector, we can fit the
most obvious stellar sequences, which has led us to identify three distinct stellar populations.
This displacement is illustrated by dotted lines (in black) 
for  three spectral type along the reddening direction. This permits to define
the spectral type of reddened stars.\\
In what follows we discuss how we have chosen these amounts and why we claim we are distinguishing
three different populations.\\

\noindent
The yellow dot-dashed ZAMS has been shifted by $E(B-V)$ = 0.25, and fits a conspicuous sequence
of early type stars.  As indicated in the figure, this group is composed by stars of spectral
type from as early as O8 up to A0.  Beyond A0, the definition of spectral type becomes ambiguous
due to the crossing of different reddening  ZAMS close to the A0V knee. This low-reddening
group  of stars clearly suffers from variable extinction, as indicated by the significant
spread around the  $E(B-V)$ = 0.25 ZAMS.
We estimate in fact that the mean reddening
for this group is 0.25$\pm$0.10. Sh04 identified this group as stars belonging to the Carina
branch of the Carina-Sagittarius spiral arm.  We will refer to this group of stars as group
{\bf A}.\\

\noindent
The blue short-dashed ZAMS has been shifted by $E(B-V)$ = 0.52.  It passes through a group of
young stars, with spectral types ranging from B2 to A0, which exhibit almost no reddening
variation. In fact, we estimate that the mean reddening for this group is 0.52$\pm$0.05, namely
the scatter around the mean value of 0.52 is small, and compatible with the typical photometric
error in the (U-B) and (B-V) colors.
In Fig.~3 of
Sh04 this group is barely visible, and the authors make no comments about it. Because of the
wider area covered by our study, it could be readily detected in our $(B-V)$ vs. $(U-B)$
color-color diagram.  We will refer to this group of stars as group {\bf B}.\\

Similar features (namely secondary sequences in two-color diagrams) have been found in a
variety of two-color diagrams of stellar clusters and field stars in the third Galactic
Quadrant (see e.g. Carraro et al. 2005). They are though to be produced by early type stars stars belonging
to  distant spiral features in the outer Galactic disk (see also Moitinho et al. 2008).\\

\noindent
The two red long-dash ZAMSs have been shifted by $E(B-V)$ = 1.10 and $E(B-V)$=1.50, 
and identify a group of
early-type (from O6 to B5) stars suffering strong reddening with a significant dispersion.
Since the region in the color-color diagram between the two red ZAMS is continuously occupied by stars,
we propose they basically belong to the same distant population.

We estimate that the mean reddening for this group is 1.30$\pm$0.20. 
Within this group of stars, 
Sh04 identify a cluster of young stars
surrounding the two WR stars (the two crosses in Fig.~5). 
We will refer to this group of stars as group {\bf C} or
Shorlin~1 (also known as C1104-610a).\\

\noindent
As anticipated, the two crosses in Fig.~5 identify the two WR stars; WR38 (WC4) and WR38a (WN5). As exhaustively
discussed by Sh04, $U$-band photometry of this type of star is difficult, but they do provide
corrections to transform their colors to the Johnson-Morgan standard system.  Given that our
uncorrected colors (see Table~2) are in good agreement with theirs, we adopted their corrections
(see the Table~3 in  Sh04).

\begin{table}
\tabcolsep 0.1truecm
\caption{Photometry of the stars WR38 and WR38a}
\begin{tabular}{lccc}
\hline
\noalign{\smallskip}
& V & (B-V) & (U-B) \\
\noalign{\smallskip}
\hline
\noalign{\smallskip}
WR38  & 14.684 &  1.216  & 0.679\\
WR38a & 15.113 &  1.191  & 0.077\\
\noalign{\smallskip}
\hline
\end{tabular}
\end{table}

\section{Properties of groups A, B and C}

To be more quantitative, we first measured the individual reddening of the stars in groups {\bf A}, 
{\bf B} and {\bf C} using the well known reddening free $Q-$ parameter (see for instance Johnson 1966), 
which allows to establish the membership of a certain
star to a group on the basis of common reddening (see e.g.  V\'azquez et al. 1996,  V\'azquez et al. 2005, and  Carraro 2002).
Having provided evidences that the reddening law is normal, the Q-parameter is therefore defined as:

\[
Q = (U-B) - 0.72 \times (B-V)
\]

\noindent
The value of Q is a function of spectral type and absolute magnitude (Schmidt-Kaler 1982).
\\

This technique permits to identify {\it bona fide} members of similar spectral type as late as
A0V. We have identified 146 stars in our sample with a mean $E(B-V)$ of 0.25$\pm$0.10 (group {\bf A}),
96 stars with a mean $E(B-V)$ of 0.52$\pm$0.05 (group {\bf B}), and 212 stars with a mean $E(B-V)$
of 1.30$\pm$0.20 (group {\bf C}).\\

\noindent
Then, to obtain an estimate of the mean distance of the three groups, we made use of the
variable extinction diagram shown in Fig.~6, where to derive  ($V-M_V$) we 
adopt M$_V$ as a function of sprectral type from Schmidt-Kaler 1982.  
Yellow triangles, blue squares and red circles
depict members of groups {\bf A}, {\bf B}, and  {\bf C}, respectively. The four lines plotted
have been drawn adopting the normal extinction law ($R_V$ = 3.1), and correspond to four different
absolute distance moduli (V$_{0}$ - M$_{V}$): 10.0 (solid), 12.0 (dotted), 14.0 (short-dashed) and
16.0 (long-dashed). The histograms in the right panel of this figure give the apparent distance
moduli distribution, from which we infer the mean distance to the three groups.\\ 
One can in fact extrapolate the absolute
distance modulus (m-M)$_{V,o}$ reading the value along the $Y$-axis, where the reddening is zero.\\

   \begin{figure}
   \centering
   \includegraphics[width=9.0truecm]{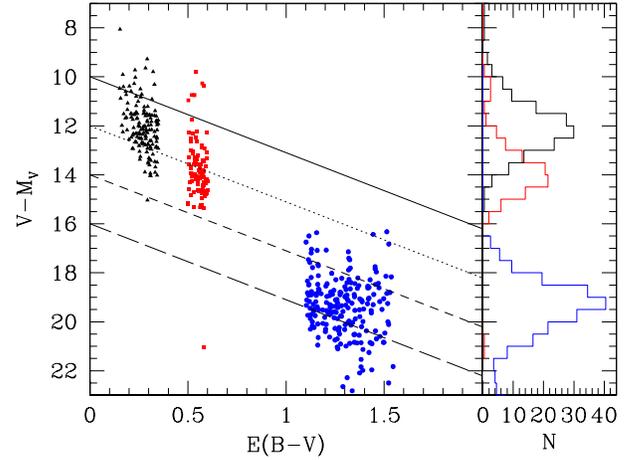}
   \caption{Variable extinction diagram for all early type star for which individual reddening
    was  measured.  The four lines plotted have been drawn adopting the normal extinction law
    ($R_V$ = 3.1), and correspond to four different absolute distance moduli (V$_{0}$ - M$_{V}$):
    10.0 (solid), 12.0 (dotted), 14.0 (short-dash) and 16.0 (long-dash). The histograms in the
    right panel of this figure give the apparent distance moduli distribution, from which we
    infer the mean distance of the three groups.}%
    \end{figure}

\noindent
In the left panels of figures 7, 8 and 9 we present $(B-V)$ vs. $(U-B)$ color-color diagrams 
enhancing the stars of groups {\bf A} (yellow triangles), {\bf B} (red squares) and {\bf C}
(blue circles), respectively.
The right panels of these three figures show the distribution on the plane of the sky for the objects
in each group. In these latter panels stars have been plotted as circles scaled according to their
$V$ magnitude. The fields shown in these figures are centered at RA = $11^{h}05^{m}48.5^{sec}$, 
DEC = $-61^{o}11^{\prime}5.9^{\prime \prime}$.
North is up, East to the left as in Fig.~1.\\

   \begin{figure}
   \centering
   \includegraphics[width=\columnwidth]{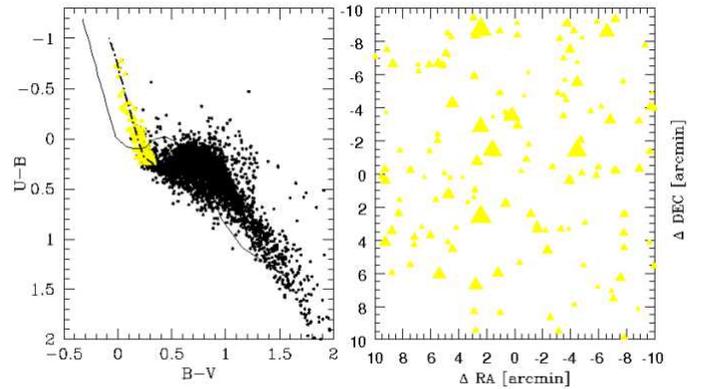}
   \caption{Stars belonging to group {\bf A} (yellow triangles).  Left panel (as well as left
   panel of Fig.~6) illustrates how this group was selected on a reddening basis. See text for
   details.}%
   \end{figure}

   \begin{figure}
   \centering
   \includegraphics[width=\columnwidth]{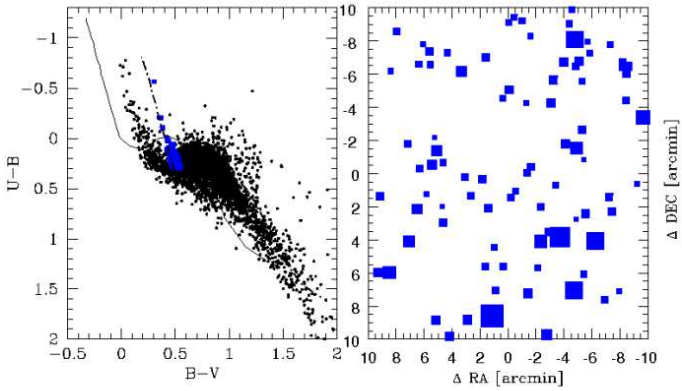}
   \caption{Same as Fig.~6, for stars belonging to group {\bf B} (red squares). See text for
   details.}%
   \end{figure}

   \begin{figure}
   \centering
   \includegraphics[width=\columnwidth]{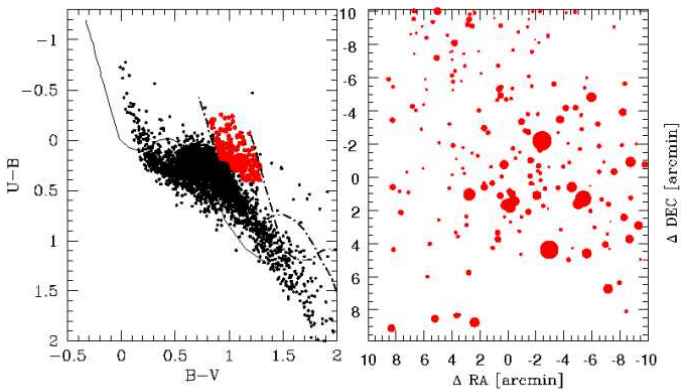}
   \caption{Same as Fig.~6, for stars belonging to group {\bf C} (blue circles). See text for
   details.}%
   \end{figure}

\subsection{Group A}

Inspection of the right panel of Fig.~7 shows that the objects in group {\bf A} are evenly distributed
on the sky, with no hints for any particular concentration.\\

The sequence defined by this group in variable reddening diagram (Fig.~6) has a mean intrinsic
distance modulus of 12.0 magnitudes, with a large spread. Still, most of the stars lie at a mean
distance of 2.5$\pm$1.5 kpc. This mean distance is compatible with this group lying in the part
of the Carina branch of the Carina-Sagittarius arm closest to the Sun. The observed distance spread
is compatible with the typical size of inner Galactic spiral arms (1.5-2.0 kpc, Bronfman et al. 2000).\\

We therefore conclude that group {\bf A} is composed of young field stars belonging to the Carina-Sagittarius
arm, confirming the suggestion by Sh04.  We dispute however the possibility that these stars form a
star cluster as suggested by Shorlin (1998), and claim that this conclusion was possibly the result of
the small FOV covered by the photometry of Shorlin (1998). Our wider coverage clearly shows that we
are looking at a population of field stars having the same reddening and age, but located at different
distances in the Carina arm.

\subsection {Group B}

As was the case of group {\bf A}, the objects in group {\bf B} are also evenly distributed on the
plane of the sky, but they are, on the average, fainter (see right panel of Fig.~8).\\

In the variable reddening diagram (Fig.~6) they trace a sequence of stars which lie at larger distance  than
those of group {\bf A}.  They have a mean absolute distance modulus of 14.0 magnitudes, which implies
a bulk distance of 6.0$\pm$2.0 kpc. The  stars in this group are also young, and, according to modern
descriptions of the MW spiral structure (Vall\'ee 2005, Russeil 2003) they are most probably 
located in the part of the Carina branch more distant from the Sun. In fact, the tangent to the
Carina-Sagittarius arm points to $l\sim280^o$, in a way that the line of sight to these groups is
expected to cross twice the Carina portion of this arm.\\

This group had not been noticed before, which may be the origin of the different conclusions
suggested by Sh04 and Wa05 on the location, and distance, of the putative star cluster associated to
WR38 and WR38a (see discussion below).

\subsection {Group C}

This is the most interesting group of faint young stars with common reddening. Our mean $E(B-V)$
value (1.30$\pm$0.20) is smaller than the value found by Sh04, but still marginally compatible
within the errors declared. The 212 stars extracted within this reddening range have a broad distance
distribution (see Fig.~6), are faint, and fairly evenly distributed across the field of view
(see Fig.~9). Careful inspection of Fig.~9 indicates the existence of a concentration of stars close
to the center of the field. This concentration corresponds to Shorlin~1 (C1104-610a, Sh04).\\

\noindent
The point now is whether this is really a cluster  or just a chance concentration of bright stars-
and where it lies exactly.  As Sh04 argue, the reality of the cluster can be assessed only with much
deeper photometry. To address these two issues we will make use of the variable reddening diagram 
(Fig.~6) and of the CMDs presented in Fig~10 and Fig.~11, constructed with the objects belonging to
group {\bf C} having the largest reddening (E(B-V) $>$ 1.45), as Shorlin~1 (see Sh04). 
Filled symbols in Figs.~10 and 11 indicate stars within a $0.6\arcmin$ wide circle
centered on Shorlin~1 (nominal center RA = $11^{h}05^{m}46.52^{sec}$; DEC=$-61^{o}13^{\prime}49.1^{\prime\prime}$), 
which is the zone where the cluster seems to be located.\\

   \begin{figure}
   \centering
   \includegraphics[width=\columnwidth]{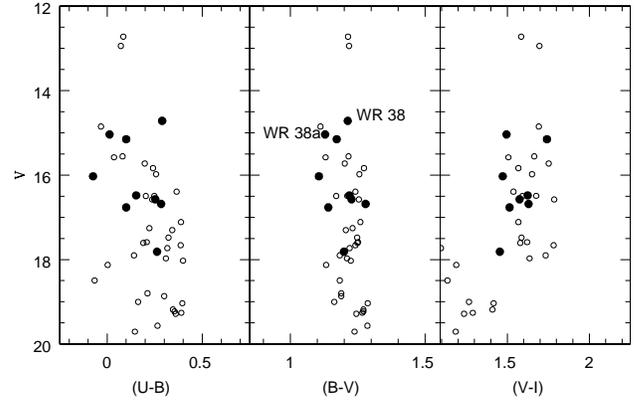}
   \caption{Color-Magnitude Diagrams for stars belonging to group {\bf C}, for different color
   combinations. Filled symbols indicate stars inside the Shorlin~1 area}%
   \end{figure}

   \begin{figure}
   \centering
   \includegraphics[width=\columnwidth]{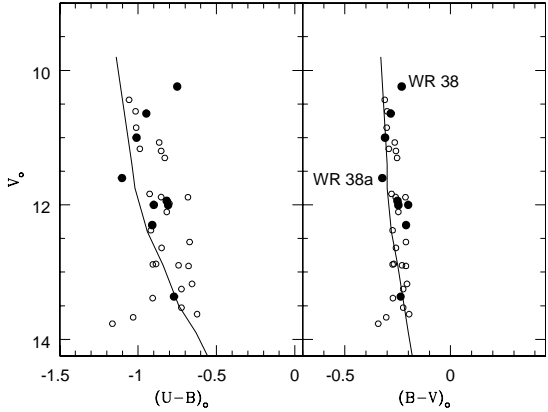}
   \caption{Reddening free Color-Magnitude Diagram for stars belonging to group {\bf C}, for
   different color combinations. Filled symbols indicate stars inside Shorlin~1 area. In this figure only
   stars having E(B-V) larger than 1.45 are plotted. See text for more details.}%
   \end{figure}

\noindent
Given that these common reddening stars of group {\bf C} distribute evenly in the field (see Fig.~9),
we can compare the position (in a CMD) of objects in  the general field with that of objects in
the Shorlin~1 region. As shown by the CMDs presented in Fig.~10, general field stars (open symbols)
occupy the same region as Shorlin~1 stars (filled symbols). This suggests that they share the same
distance range and age, and therefore conform a common properties, evenly distributed, population.\\

The same conclusion can be drawn from the reddening-free CMDs presented in Fig.~11. In these
plots the stars within the putative cluster area and those in the general field define the same
sequence.  The $V_o$ vs. $(B-V)_o$ CMD presented in the right panel of Fig.~11 has been plotted
in the same scale as Fig.~5 of Sho04. Note that, in spite of our photometry being six magnitudes
deeper in $V$, we detect only one fainter object in the cluster area. We can be confident therefore
that there are no fainter stars associated to the cluster.\\

The ZAMS superposed in the both panels of Fig.~11 is for (m-M)$_{V,o}$ = 15.5, which suggests that
the distance of group {\bf C} is about 12.7$\pm$3.0 kpc. Both Shorlin~1 members and field stars exhibit
however a significant distance spread (see Fig.~6)\\

For the reasons summarized below, we conclude that there is no clear evidence of a star
cluster at the location of WR38/WR38a.

\begin{description}

\item $\bullet$ a cluster hosting two WR stars would have to be a relatively massive one, with
numerous stars of spectral type later than B; on the contrary to what we see,

\item $\bullet$ we do not see any special concentration of stars at the putative cluster location
(see Fig.~1),

\item $\bullet$ the sequence defined by Shorlin~1 in the CMDs is also occupied by field stars which
have the same reddening, and are evenly distributed in the field (see Figs. 9, 10 and 11),

\item $\bullet$ the stars defining the cluster (the 8 stars discovered by Sh04, plus one fainter
object from this work) exhibit a significant distance spread (see Fig.~6), incompatible with a
physical, although loose, star cluster.

\end{description}

\noindent
To summarize our findings, we report in Table~3 the main properties of the three
groups. As for the age we only provide an upper limit, since the populations under
investigations are surely young, and a precise age estimate - by the way quite difficutl- 
is beyond the scope of this work.

\section{Conclusions}

The large angular coverage, and depth, of our $UBVI$ photometry of the region around the
WR38/WR38a has allowed us to clarify the stellar populations towards Galactic longitude
$\approx$ 290$^o$, in the fourth Galactic quadrant.\\

\noindent
Our results can be visualized with the aid of Fig.~12, where the spiral structure of the
Galaxy (Vall\'ee 2005) is presented in the $X-Y$ plane, where $X$ points in the direction of
Galactic rotation and $Y$ points towards the anti-center.  The Carina and Perseus arms are
indicated with the symbols $I$ and $II$, respectively. The Galactic center and the position
of the Sun are also indicated, at (0.0,0.0) and (0.0,8.5), respectively. The position, pitch
angle, and extension of the arms are clearly model dependent; to a lesser extent for the Carina arm
(which is well known) and significantly more for the less-known Perseus arm (both HI observations
(Levine et al. 2006) and HII observations (Russeil 2003) coincide however on the approximate
location and extent of the Perseus arm).  Beyond  $\sim$9 kpc from the Sun, it is not expected 
to find spiral features related to the Carina arm (Georgelin et al. 2000); more distant structures
can be associated with the Perseus arm. According to Levine et al. (2006, their Fig.~4) at l=290$^o$
the Perseus arm is located in the range $ -18.0\leq X \leq -13.0 $ and $0.0 \leq Y \leq 2.0$ Kpc,
respectively, in nice agreement with Vall\'ee model.\\ 

\noindent
With a dashed arrow we indicate the line-of-sight in the direction of our field, and with thick
segments the distance range we derived for the three groups ({\bf A}, {\bf B} and {\bf C}) we
identified in this Galactic direction (see also Table~3).  Fig.~12  shows that the groups detected remarkably fit
the position of the Carina and Perseus arm. This confirms Sh04's suggestion that the most distant
group ({\bf C}) is most probably associated with the extension of the Perseus arm in the fourth 
Galactic quadrant, and constitutes the first optical detection of this arm. However, on the
contrary to Sh04, here we propose that this extreme group is not a star cluster, but simply a group
of young stars uniformly distributed within the Perseus arm.\\

The direction we are investigating is about $\sim1^o$ below the Galactic plane. At distances of
2.5, 6.0 and 12.7 kpc this implies heights below the Galactic plane of about 50, 100, and 210 pc,
respectively.  These are sizeable values, and reflect the trend of the Galactic warp in this 
zone of the disk (Momany et al. 2006).\\

   \begin{figure}
   \centering
   \includegraphics[width=\columnwidth]{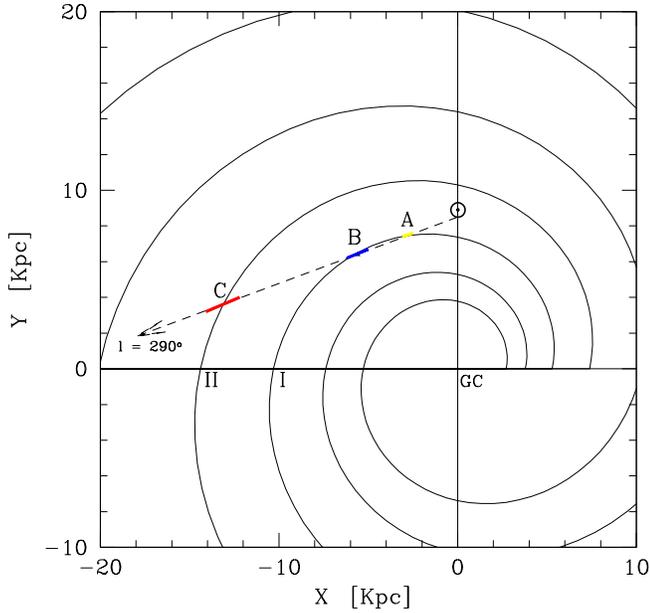}
   \caption{Schematic view of the spiral structure of the Galaxy (Vall\'ee 2005). $X$ points in
   the direction of Galactic rotation, and $Y$ points towards the anti-center. The position of the Sun
   (0.0,8.5) and of the Galactic center (0.0,0.0) are indicated. The symbols I and II indicate the
   Carina and Perseus spiral arms, respectively. Thick segments illustrate the location of the three
   groups ({\bf A}, {\bf B} and {\bf C}) discussed in the text. See text for details}%
   \end{figure}

\begin{table}
\tabcolsep 0.1truecm
\caption{Properties of the three distinct groups we detected in the field toward WR~38 and WR~38a.}
\begin{tabular}{lcccc}
\hline
\noalign{\smallskip}
Group& E(B-V) & d$_{\odot}$ & age & Note\\
\noalign{\smallskip}
\hline
\noalign{\smallskip}
A & 0.25$\pm$0.10     &  2.5$\pm$1.5  & $<$ 100 Myr& Carina\\
B & 0.52$\pm$0.05     &  6.0$\pm$2.0  & $<$ 100 Myr& Carina\\
C & 1.30$\pm$0.20     & 12.7$\pm$3.0  & $<$ 100 Myr& Perseus\\
\noalign{\smallskip}
\hline
\end{tabular}
\end{table}

\noindent
We finally wish to discuss our photometric database in the framework of a possible detection
of the Argus over-density discussed in Rocha-Pinto et al. (2006), or the Monoceros Ring, 
which Shorlin~1 was suggested to be associated with (Frinchaboy et al. 2004). With this aim, in
Fig.~13 we present $V$ vs. $(B-V)$ CMD  for all the stars having BV  photometry.\\ 

   \begin{figure}
   \centering
   \includegraphics[width=9.5truecm]{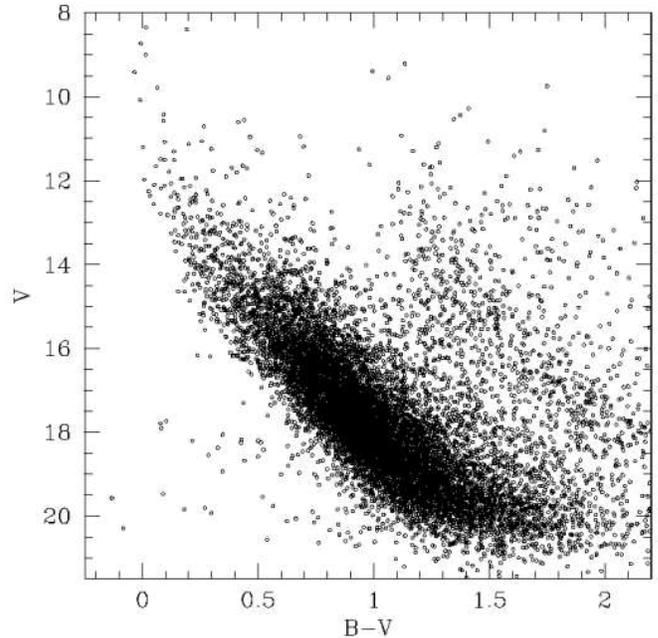}
   \caption{CMD of all stars with BV photometry in the field under study}%
   \end{figure}

The distribution of stars in this diagram is relatively easy to describe. Apart from the
prominent blue sequence of young stars brighter than V$\approx$ 14.0, which are the objects of
{\bf A}, we can distinguish two other remarkable features. The first is a thick Main Sequence (MS)
downward of V$\approx$ 14.0, widened by photometric errors and binaries, but mostly reflecting the
fact we are sampling objects at very different distances, and with different reddening, in the
Galactic disk. Along this thick MS we do not find any evidence of a blue Turn Off Point, typical
of the metal poor, intermediate age, population of the Monoceros Ring (Conn et al 2007).\\
Therefore we conclude that no indication of the presence of the Monoceros
Ring or the Argus system is detected in our field.\\
\noindent
The second is a Red Giant Branch, significantly widened and somewhat bent by variable reddening, 
composed of giant stars at different distances. The relatively high number of giants can be easily accounted for,
because we are sampling a field located 1$^o$ below the Galactic plane.\\

\noindent
Our results confirm the effectiveness of multicolor optical photometry in the study of the 
structure of the MW disk.  More fields need to be observed to better constrain the
spiral structure in the fourth Galactic quadrant, in particular the shape and extent of the
Perseus arm, and, possibly, detect the more distant Norma-Cygnus arm.\\
At odds with radio observations, which has to rely on the poorly known Galactic rotation
curve, optical observation, especially in low absorption directions, 
can better constraint the distance to spiral
features.

\begin{acknowledgements}
Tomer Tal and Jeff Kenney are deeply thanked for securing part of the observations
used in this paper. GC acknowledges R.A.V\'azquez for very
fruitful discussions. EC acknowledges the Chilean Centro de Astrof\'isica FONDAP
(No. 15010003).
\end{acknowledgements}

\end{document}